# Towards the automation of book typesetting

Sérgio M. Rebelo[a*], Tiago Martins[a*], Diogo Ferreira[a*], Artur Rebelo[a]

[a] University of Coimbra, Centre for Informatics and Systems of the University of Coimbra, Department of Informatics Engineering, Coimbra, Portugal

**Abstract**

This paper proposes a generative approach for the automatic typesetting of books in desktop publishing. The presented system consists in a computer script that operates inside a widely used design software tool and implements a generative process based on several typographic rules, styles and principles which have been identified in the literature. The performance of the proposed system is tested through an experiment which included the evaluation of its outputs with people. The results reveal the ability of the system to consistently create varied book designs from the same input content as well as visually coherent book designs with different contents while complying with fundamental typographic principles.

**Keywords**

Design tools; Data-driven Design; Generative Design; Graphic design; Typography.

[*] Corresponding authors.
Email addresses: srebelo@dei.uc.pt (Sérgio M. Rebelo), tiagofm@dei.uc.pt (Tiago Martins), df@diogoferreira.com (Diogo Ferreira)

# 1. Introduction

Typography is the art of giving our language a visual form. Thus, it is through typography that we materialise and store our knowledge and information [1]. Since the publication of the first typographic book, in the mid-fifteenth century, society has been looking for the most appropriate way to convey a message typographically. In the field of editorial design, this effort has focused on the search for the best principles for designing typographic compositions, such as books [2].

Advances in digital technologies have been changing the work process of graphic designers. The emergence of computational approaches in the design domain enabled designers to explore new perspectives, new conceptual and visual possibilities, and achieve new types of solutions. Furthermore, the emergence of computational approaches comes with a paradigm shift in the role of the designer, who begins to create processes that enable the creation of designs, instead of designing the final solution. In other words, the design concept is translated into a computer program that systematically explores various design possibilities from the original concept. That said, in the particular case of layout design, we consider that the potential of computational approaches is not yet being fully explored.

In this work, we explore a computational generative approach for the automatic design of book layouts. The result is a computer system, which operates inside the Adobe InDesign environment, that automatically generates book designs from input content. Figure 1 shows different books created with the presented system. It starts by receiving the input content, namely text and images. Before generating compositions for the input content, the designer can specify restrictions on some of the visual characteristics of the output compositions, *e.g.* format, size, grid and font. Then, the system automatically typesets the content based on a set of typography rules and principles found in fundamental literature in the field. In the end, the system presents the generated composition to the user as an editable Adobe InDesign document.

Overall, the system is capable of creating layout compositions that comply with specific fundamental typographic principles while matching the graphic preferences of the user. In addition, experiments conducted with the system demonstrate its ability to autonomously generate varied compositions with the same input content and also generate visually coherent compositions for different input contents. In addition, it creates functional layout designs in an almost unpredictable manner. This reveals the great potential of this approach to layout design, both in generating outputs that the designer uses as final solutions and in using the outputs as starting points for further explorations. This work is aligned with the experiments previously developed by Ferreira *et al.* [3]

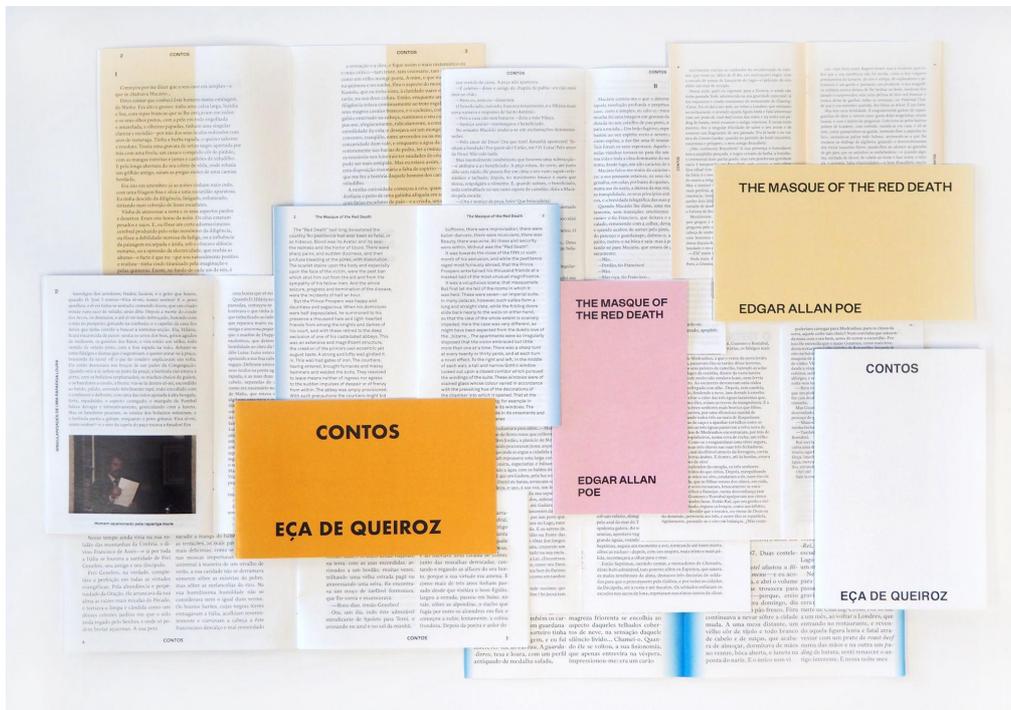

**Figure 1.** Examples of books generated with the presented system. All designs generated by the system for this paper, along with demonstration videos, can be found in the supplementary files.

The remaining of the article is organised as follows: Section 2 overviews related works; Section 3 describes the system, namely the interaction process, the engine that runs the system, as well as its inputs and outputs; Section 4 presents an experiment conducted to validate the designs created with the system and discusses the results obtained; Lastly, Section 5 summarises the main contributions of this work and identifies future work.

## 2. Computational Approaches in Editorial Design

Systematic approaches have been popular in layout design since the mid-twentieth century when some creative practitioners designed layouts based on grids and the variation of the visual features of typographic elements [4–6]. Some works have explored algorithmic approaches for book layout design. In the 1960s, Gerstner introduced a selective and combinatorial method for the design of graphics, including layouts [7]. Afterwards, he translated it into a logical language that computers could understand in *Compendium for Literates: A system of Writing* [8]. LeWitt compiled, in 1971, a set of formal instructions to design a conceptual art exhibition catalogue [9]. Already in the 1980s, Knuth and Plass [10] presented a dynamic programming algorithm to page breaks avoiding widows and orphans and employed it to typeset a two-column dictionary. Soon thereafter, Knuth introduced the parametric typeface design language, *Metafont* [11] and the *TeX* typesetting system [12], enabling anybody to produce a book using a structural markup language and a set of high-level commands. Cooper and her students, at the Visible Language Workshop, experimented with the generation of layouts, *e.g.* publications that resulted from the collaboration with IBM [13] and the cover for *Design Quarterly 142* [14]. In the mid-1990s, Maeda also designed a series of digital booklets, the *Reactive Book* series, where the graphics are controlled interactively by user input [15].

Nevertheless, in the last two decades, we observed the increasing employment of the use of computational graphic design approaches, especially because of the development of easier-to-use creative code environments [16,17]. A solid overview of the field is presented, for instance, by Reas *et al.* [6] or Richardson [18]. These approaches have been explored in several artistic and creative domains, including the generation of visual and communication artefacts such as poster designs (*e.g.* Rebelo *et al.* [19] or Guo *et al.* [20]), banners (*e.g.* Gatarski [21] or Yin *et al.* [22]), user interfaces (*e.g.* Quiroz *et al.* [23] or Amitani *et al.* [24]), visual identities (*e.g.* Levin [25] or Neue [26]), type designs (*e.g.* Ahn & Jin [27] or Martins *et al.* [28]), among others. In the context of book typesetting, two frameworks are stimulating the adoption of computational practices. The library *Basil.js* [29] provides friendly and accessible tools for scripting and automation in Adobe InDesign. Then, the open-source framework *The Magic Book* [30] facilitates the design, production, and self-publishing of books.

Tailor-made procedural and template-based approaches are employed in the generation and definition of layouts, modification of layouts based on their contents and/or inter-relationships between the elements. LettError type design studio used random processes and parametric design methods to design several typographic artefacts such as calendars, type specimens and even their portfolio [31,32]. Oliveira [33] presented a recursive division method to place elements on both one and multiple-column grid layouts. Cleveland [34] proposed a method for generating style-based design layouts that explores the inter-relationships between text and graphics. Also, he presented a system to generate layouts employing these principles. LUST developed a set of scripts to layout and stylise the book "I Read Where I am" informed by its content [35]. Damera-Venkata *et al.* [36] presented a template-based probabilistic framework for generating document layouts for variable content. Ahmadullin and Damera-Venkata [37] also presented a probabilistic model for newspaper typeset that, based on given content, divides the available layout into regions and optimises the content to fit within these regions. Flipboard developed *Duplo* [38], a layout engine that creates news magazines adapted to its contents and based on a set of heuristics such as the amount and flux of text or the existence and position of images.

We can also observe the use of Artificial Intelligence approaches for typesetting. Evolutionary computation and greedy approaches have been used to create layouts with varied purposes. For example, Geigel and Loui [39] evolved layouts for photo books by evaluating different aesthetic criteria. Goldenberg [40] employed an evolutionary approach to automatically generate page layouts, minimising the waste of space on the page. Gonzalvez *et al.* [41] used a greedy simulated annealing algorithm to create multi-column newspaper layouts. Purvis *et al.* [42] automatically evolved documents using a multi-objective optimisation approach, considering a set of layout constraints and aesthetic measures. Quiroz *et al.* [43] evolved brochure documents according to user preferences and design guidelines. Strecker and Hennig [44] proposed a grid-based method for newspaper layouts, minimising the wasted space and bearing in mind newspaper design aesthetic measures. Boll *et al.* [45] and Sandhaus *et al.* [46,47] evolved photo layouts based on rules of layout design and proposed a method to transform a blog into a photo book considering different aesthetic requirements. Önduygu [48] developed the system *Gráphagos* that generates compositions through the interactive evolution of specific features of visual elements. Klein [49] developed the tool *Crossing, Mixing, Mutating* to create variations in a template using genetic operators. Later, an updated version of this tool was released as an InDesign plug-in named *Evolving Layout* [50]. Lopes *et al.* [51] developed the system *evoDesigner*, which automatically creates and evolves designs in the InDesign environment. Recently, Machine Learning approaches have been employed in the layout design field taking into account the relation between elements on layout and the learning of specific typesetting and design styles, *e.g.* Zheng *et al.* [52], Li *et al.* [53], or Kikuchi *et al.* [54].

Our analysis of the related work unveils a series of computational approaches that reveal great potential for the support and automatisation of the creation of book layouts. However, as far as we know, the existing approaches on this context rarely enable the subsequent modification of the generated designs, even less in the natural working environment of the designer. This way, it is difficult to introduce such computational technologies in the design process in an easy and fluid fashion. Also, we denoted that most of these approaches do not present a multipurpose objective, being developed to generate specific designs.

## 3. System

We developed a computer system to automatically typeset books from content provided by the user. This system is developed as a computer script that operates inside Adobe InDesign software which is popular among graphic designers working in the field of typography. We idealised and implemented the system to take advantage of the typeset functionalities supplied with InDesign, which can be controlled via scripting. By integrating the system with InDesign, we allow users to generate design variations and easily edit them within a familiar working environment. One can find demonstration videos in the supplementary files. To install the system in the InDesign environment, the user only needs to copy the folder that contains the system files to the scripts folder in the application directory. To facilitate access to the system, we also made available a script that creates a dedicated tab for the system in the InDesign navigation menu. The system installation files and source code are available at https://cdv.dei.uc.pt/2019/scriptedpages.[1]

In the following subsections, we describe the user interface of the system and explain how its engine works.

### 2.1 Interface

Figure 2 shows different snapshots of the user interface built to enable the interactive control of the inner workings of the system. The user interface is structured in a series of five tabs. In each tab, one can set specific properties of the composition or let the system choose automatically based on a set of predefined rule-based values for those properties.

The first interface tab "Document" (Figure 2a) concerns the structural characteristics of the document. It allows the user to set the page size, margins, number of columns and gutter. There is also an option to import settings stored in a file.

---

[1] We also made available supplementary materials, including demonstration videos and examples of books designed with the proposed system.

In the second tab "Input" (Figure 2b), the user provides the content of the book to be typeset. To that end, the interface offers two options. The first option is to load a Microsoft Word file containing only text, text and images, or only images. The second option is to load a Microsoft Word file containing only the text and a folder containing the images. For this second option, the place where each image should be inserted must be identified in the text using a tag @imageFileName@, which will be replaced by the image with the same name contained in the loaded folder. In addition, the user can choose whether to generate a table of contents and/or colophon for the book. The last option of this tab allows the user to select the language of the input text so that it is possible to correctly hyphenate words.

The third tab "Styles" (Figure 2c) concerns the definition and mapping of paragraph styles. The user has three options to choose from. The first option is to keep the styles imported from the input Word document. The second option is to map each style of the input document to another paragraph style selected, manually or randomly, from a list created from all fonts installed on the computer, while keeping the remaining paragraph attributes. The last option is to let the system generate the styles, suggesting font combinations based on the rules entered in the system (these rules are explained in the next subsection).

In the fourth tab "Experimental" (Figure 2d), the user can toggle experimental features that can be applied to the generated book. In the presented version of the system, there are four experimental features: (i) draw a colour background on half of each book page; (ii) draw a colour gradient along the inner and/or outer margins of the pages; (iii) apply a random indentation to each text paragraph; and (iv) make the cover title as large as possible. The purpose of these features is to increase the uniqueness of the resulting designs. The user can also opt to let the system randomly choose if any experimental features should be applied by selecting the checkbox "Surprise me." Furthermore, new features can be implemented and added to the system at any time.

After interacting with these four tabs of settings, the user can start creating book designs by clicking on the button "Create" located at the bottom right corner of the interface. This button will start the engine of the system which will automatically create a new InDesign document and typeset a new book from the content and settings chosen by the user.

Once the typeset process is complete, the result is presented to the designer as an editable InDesign document. From that moment on, the user can, for instance: (i) adopt the generated book as a final design; (ii) use the generated book as a starting design from which the designer can make any changes or refinements needed; or (iii) continue to use the system to generate more designs until a more suitable design is found.

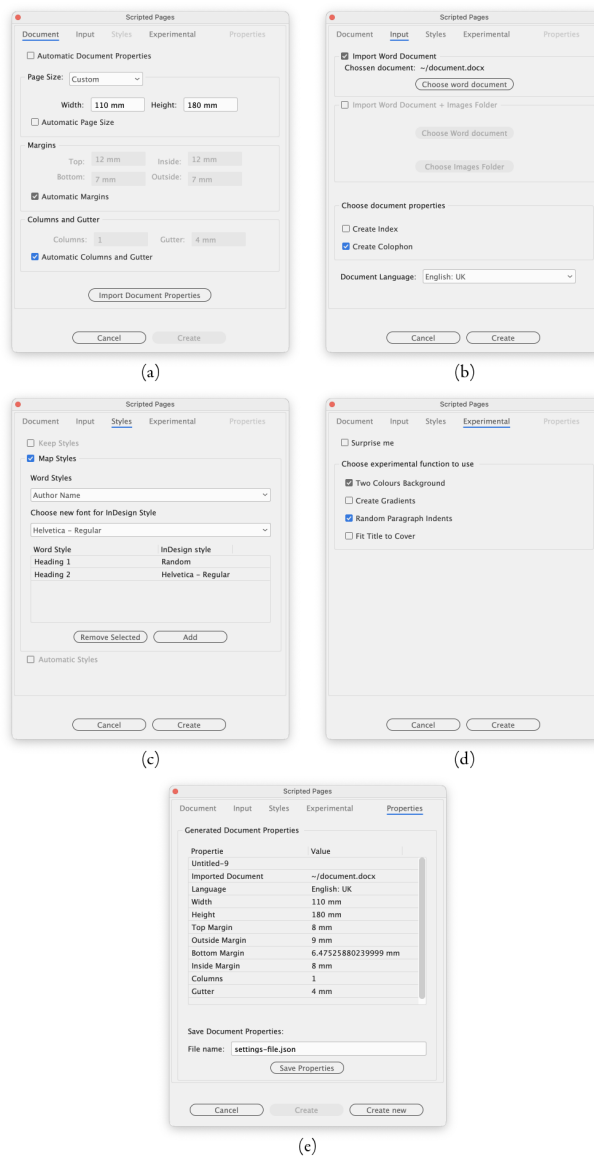

**Figure 2.** Snapshots of the system, showing the five different tabs of the user interface: (a) Document, (b) Input, (c) Styles, (d) Experimental and (e) Properties. Demonstration videos of the system can be found in the supplementary files.

There is another interface tab, entitled "Properties" (Figure 2e), which not only overviews the settings used in the generation of a book but also enables the user to save those settings to a file. Later, this settings file can be imported to the system using the first interface tab, as mentioned earlier. This functionality can be useful, for example, to facilitate the typeset of different books using the same settings. This last tab is only accessible after a book is generated. After the generation, it is also made visible a button that enables the user to generate a new book, maintaining the same input document and predefined settings.

## 2.2 Engine

The system operates based on a series of typographic rules, styles and principles (Table 1) which were identified and collected from literature recognised in the field. This includes the work by Bringhurst [1], Müller-Brockmann [55], Haslam [56], Hochuli and Kinross [2], Hochuli [57], Lupton [58] and Tschichold [59]. The encoding of these rules into the system, enables the typeset process to go through them, one by one, and make typographic decisions on all composition attributes. The rules are stored in a JSON file, which contains the possible configurations (*i.e.* values or range of values) and, when applicable, their probability rate for each typographic attribute. This file enables easy access to all rules by the system and their quick editing by the designer at any time.

**Table 1** Typography rules, styles and principles loaded into the system by default

| Attribute | Valid approaches or values | Influenced by | Based on |
|---|---|---|---|
| **Book type** | Short reading (<50,000 words) | n/a | Defined empirically based on examples |
| | Long reading (≥50,000 words) | | |
| | Only images | | |
| | Text and images | | |
| **Book Size and Format** | 105 × 180 mm (portrait) | Book type | Defined empirically based on examples |
| | 110 × 170 mm (portrait) | | |
| | 110 × 180 mm (portrait) | | |
| | 110 × 220 mm (portrait) | | |
| | 130 × 200 mm (portrait) | | |
| | 150 × 210 mm (portrait) | | |
| | 170 × 240 mm (portrait) | | |
| | 180 × 180 mm (square) | | |
| | 200 × 110 mm (landscape) | | |
| | 200 × 120 mm (landscape) | | |
| | 230 × 120 mm (landscape) | | |
| **Margins** | Defined randomly based on a certain range for each margin. Top and Bottom margins between 7 mm and 15 mm. Inside and outside margins between 7 mm and 30mm. | n/a | Defined empirically based on examples |
| **Grid** | Defined based on a random column size value calculated based on a certain range. Column sizes vary between 70 mm and 140 mm. | Book size | Müller-Brockmann [55] |
| **Line length** | Between 45 and 75 characters (the ideal is 66); ≥48 characters for justified text | Grid: Alignment | Bringhurst [1] |
| **Words per page/ Lines per page** | ≤500 words (≈45 lines) for one column | n/a | Bringhurst [1] |
| | ≤1.000 words for multiple columns | | |
| **Paragraph marks** | Ornaments | Book type | Bringhurst [1] |
| | Drop lines (space before the paragraph) | | |
| | Paragraph marks | | |
| | Negative indent | | |
| | Positive indent | | |
| **Alignment** | Justified for long reading | Book type | Haslam [56]; Lupton [58] |
| | Left align for long reading, titles and captions | | |
| | Right align for captions | | |
| | Centre align for titles | | |
| **Hyphenation** | Yes | Alignment (required when justified) | Bringhurst [1] |
| | No | | |
| **Leading** | Between 115% and 140% of the font size (the base value is 120%) | Font size; Font painting | Defined empirically based on examples |
| **Body font** | Serif for long reading | Book type | Bringhurst [1] |
| | Serif or sans-serif for short reading | | |
| **Title font** | Different or the same as the text, with another weight or in uppercase. | Body font | Hochuli and Kinross [2]; Hochuli [57] |
| **Font size** | Between 8 and 12 points | Book type; Font | Defined empirically based on examples |
| **Word spacing** | Between 80% and 120% (the ideal is 100%) | Required when justified | Defined empirically based on examples |
| **Letter spacing** | Between -5% and 5% (the ideal is 0%) | Required when justified | Defined empirically based on examples |
| **Running header and page numbering** | Running header on top of the page and slightly indented from the left margins. Page numbering at top of the page and aligned with the right margins. | Margins; | Defined empirically based on examples |
| | Running header on the bottom of the page and centred with text block. Page numbering at bottom of the page and aligned with outer margins (slightly indented from the text block on the left page). | | |
| | Running header on top of the page and centred with text block. Page numbering at top of the page and aligned with outer margins (slightly indented from the text block on the left page). | | |
| | Running header on top of the page and centred with text block. Page numbering at bottom of the page and centred with text block. | | |
| | Running header rotated 90º and vertically centred in the middle of the space before the outer margin. Page numbering horizontally centred at the upper corner of the outer margin. | | |
| | Running header rotated 90º and vertically centred in the middle of the space before the outer margin. Page numbering horizontally centred at the upper corner of the inner margin. | | |

Following all the encoded typographic rules, the system performs the typesetting process in order to computationally design books. This process consists of seven different sequential steps: (i) Input processing; (ii) Document size and grid definition; (iii) Typeface definition; (iv) Typographic styles definition; (v) Document typesetting; (vi) Experimental features application; and (vii) Cover design. The next subsections further describe each step.

*2.2.1 Input Processing*

The first step is to load and process the content provided by the user. Once the content is loaded, the system analyses it and extracts data concerning, for example, the text length, number of images and proportion of text in relation to images. This data is important since it allows the system to determine the type of book that it is typesetting, that is, the extracted data may indicate whether the task is to design a long reading book (*i.e.* 50,000 words or more), a short reading book (*i.e.* less than 50,000 words), a text and images book, or a book that only contains images. This information, in turn, will enable the system to make typographic decisions in the following steps of the typesetting process.

*2.2.2 Document Size and Grid Definition*

The next step is to create a new document with the page format and size based on the type of book. Thus, we encoded in the typographic styles, rules and principles a set of available book sizes as well as the probability of being selected for a certain book type. Then, the book size is selected at random based on those probability rates.

In this step, the system also defines the document grid, that is, the size of the margins, the number of columns, and the document baseline. Margins are defined randomly within a present range. The number of columns in the grid is calculated by dividing the width of the available text block by a random integer value chosen within a predefined range. Also, it considers the predefined minimum text block width. Thus, in small page sizes, it will only create one-column grids.

The template pages are defined by creating the template text boxes and placing the running header (with section name) and page numbering. The section name of a page is the last title found in the input document file. The position and size of these elements are defined empirically based on observation and analysis of examples. Currently, the system implements five different ways to compose the headers and page numbering.

*2.2.3 Typeface Definition*

Then, the system defines the typefaces to be used based on the preferences of the user, who can choose to (i) keep the typefaces used on the input document, (ii) map those typefaces to new ones installed on the computer, or (iii) let the system select the typefaces.

When users decide to map the typefaces of the input document to new ones, they must define the style for each typeface using a font installed on their computers. Typefaces that are not for mapping are defined with the same typeface as in the input document. Additionally, users can determine that the system should map one, or more, typefaces at random.

In the last option, the typefaces are selected from a set of pairs or combinations of fonts which are defined and encoded on the configuration file mentioned above. When selecting the fonts, the system considers the type of book (*e.g.* long reading or short reading). Each font pair is composed of one typeface for the titles and another for the text body, along with other typographic features that are specific to each font. Table 2 presents the typeface pairing loaded into the system by default.

**Table 2.** Typeface pairing data loaded into the system by default.

| Title font | Body font | Leading | Book type |
| --- | --- | --- | --- |
| BRRR bold (Swiss Typefaces, 2017) | PS Fournier (Typofonderie, 2012) | 1.17 | Long reading |
| Founders Grotesk bold (Klim Type Foundry, 2010) | Arnhem regular (Fred Smeijers, 2002) | 1.25 | Long reading |
| | Domain regular (Klim Type Foundry, 2013) | 1.20 | Long reading |
| | Founders Grotesk regular (Klim Type Foundry, 2010) | 1.20 | Short reading \| Text and images |
| | Tiempos Regular (Klim Type Foundry, 2010) | 1.20 | Long reading |
| Futura PT bold (Paratype, 1995) | Didot regular (Linotype, 2009) | 1.30 | Long reading \| Text and images |
| | Sabon regular (Lynotype, 1964) | 1.28 | Long reading |
| | Futura PT regular (Paratype, 1995) | 1.20 | Short reading \| Text and images \| Only images |
| Gill Sans bold (Monotype, 1928) | Baskerville PT regular (ParaType, 2016) | 1.25 | Long reading |
| | Perpetua regular (Monotype, 1925) | 1.16 | Long reading |
| | Minion regular (Adobe, 1990) | 1.28 | Long reading |
| | Times New Roman regular (Monotype, 1931) | 1.20 | Long reading |

12| | | | |
|---|---|---|---|
| GT Walsheim Pro bold (Grilli Type, 2009) | Adobe Caslon regular (Adobe, 1990) | 1.20 | Long reading |
| | Bembo regular (Monotype, 1929) | 1.20 | Long reading |
| Helvetica bold (Lynotype, 1957) | Arno regular (Adobe, 2007) | 1.16 | Long reading |
| | Joanna regular (Monotype, 1931) | 1.22 | Long reading |
| | Helvetica regular (Lynotype, 1957) | 1.20 | Short reading \| Text and images \| Only images |
| La Nord bold (Type Club Düsseldorf, 2017) | Lyon Text regular (Commercial Type, 2009) | 1.30 | Long reading |
| | Arno regular (Adobe, 2007) | 1.16 | Long reading |
| | La Nord Regular (Type Club Düsseldorf, 2017) | 1.20 | Short reading \| Only images |
| Neuzeit S bold (Linotype, 1959) | Antwerp regular (A2 Type, 2011) | 1.20 | Long reading |
| Proxima Nova bold (Mark Simonson Studio, 2005) | Arnhem regular (Fred Smeijers, 2002) | 1.25 | Long reading \| Text and images |
| FF Scala Sans bold (FontShop, 1990) | Arno regular (Adobe, 2007) | 1.16 | Long reading |
| | FF Scala Serif regular (FontShop, 1990) | 1.28 | Long reading |
| Univers bold (Linotype, 1957) | Sabon regular (Monotype, 1967) | 1.28 | Long reading |
| Akkurat bold (Lineto, 2004) | Akkurat regular (Lineto, 2004) | 1.20 | Short reading \| Text and images \| Only images |
| Antique Olive bold (Linotype, 1960) | Antique Olive regular (Linotype, 1960) | 1.20 | Short reading \| Only images |
| Arnhem bold (Fred Smeijers, 2002) | Arnhem regular (Fred Smeijers, 2002) | 1.25 | Short reading \| Text and images \| Only images |
| Fedra Sans bold (Typotheque, 2001) | Fedra Sans regular (Typotheque, 2001) | 1.20 | Short reading \| Text and images \| Only images |
| ATF Franklin Gothic bold (ATF, 2019) | ATF Franklin Gothic regular (ATF, 2019) | 1.20 | Short reading \| Text and images \| Only images |
| Scala Sans bold (FontShop, 1990) | Arno regular (Adobe, 2007) | 1.16 | Short reading \| Only images |

*2.2.4 Typographic Styles Definition*

The process continues with the definition of the paragraph, character and image styles. For each paragraph style, the following properties are defined: font, weight, size, leading, first line indentation, paragraph indentation, space before and after the paragraph, alignment, vertical alignment, hyphenation, language and colour.

The font, weight, text leading, language initial text size and colour are defined based on the preferences expressed beforehand by the user. The final text size is determined based on the chosen document grid. In this process, the system checks whether the initial typeface size is within the limits defined by the loaded rules and then it composes a text box and confirms that its median number of characters per line is within a predefined range. When the median number of characters is lower or higher than the limits, the system decreases or increases the text size, respectively. This process will continue until a proper text size is reached. When those values are achieved, if yet necessary, the document grid is also modified, namely the margins area and the number of columns. Finally, the first line indentation, paragraph indentation, space before and after the paragraph, text alignment, vertical alignment and hyphenation are defined at random based on the type of book.

The character styles are then defined based on the input document and the paragraph styles, which will be useful for using italics, bolds, small caps, among others. Also, the styles for the images are decided based on the input document and the paragraph style, thus determining their positioning, size, text wrap and effects.

*2.2.5 Document Typesetting*

Once the base document is created and all the styles are defined, the system proceeds to the actual typesetting of the book. It starts by typesetting the inside of the book. The typesetting of the inside of the book includes a sequence of steps. First, the system positions the body text. Then, the typographic styles defined earlier are applied to the entire content of the document.

The titles on the document are formatted considering three levels of hierarchies defined based on their text size on the input document: (i) chapters titles, *i.e.* titles in the biggest size and preceded by a page break; (ii) section titles, *i.e.* titles in a size bigger than the main text font and preceded by a page or a column break; and (iii) subsection titles, *i.e.* titles in the same size that the body text. Chapter titles are composed isolated on one page. Section titles are typeset on the following page of the document. In multi-column documents, they are placed isolated on the first column;

otherwise, they are placed at the beginning of the page. Subsection titles are composed inline on the text of the same size as body text.

Once the typographic styles are applied, the images and the corresponding captions are created automatically. Initially, images are placed inline, with the same width as the column where they are placed. If the book has a multi-column grid and the image is placed on the leftmost column of the document, the system randomly decides whether to change its size to fulfil more than a column. The captions are created automatically based on the name of the input images. For each document, the system determines a caption style based on the available inner space on margins as well as the existence of headers and page numbers. By default, captions can be placed below the images, aligned to the left, or aside the images, centred and vertically rotated 90º.

Finally, the table of contents and colophon are created. As mentioned before, the system interface allows the user to choose whether to create a table of contents and/or colophon. When these options are activated, a table of contents is created based on the titles on the input document and with the same paragraph style as the titles. On the other hand, the colophon with information about the generated book is typeset at the end of the book in the same style as that body text. This information includes a description of this project and parametric details about the specific design of that book, such as size, margins, and the number of columns, among others.

*2.2.6 Experimental Features*

The use of experimental features is optionally and performed when selected by the user through the interface. As mentioned earlier, new features can be developed and added to the system. The presented version of the system presents the following experimental features regarding the inside of the book. The first one draws a colour background on half of each book page, in a specific layer under the text. The second feature draws a gradient along the inner and/or outer margins of the pages. The margins where the gradient is drawn are defined randomly, being possible for the system to create gradients in both margins. In both experimental features, the background colour is defined randomly based on a set of predefined colours, as soon as the first feature is used. By default, the set of colours includes the following seven CMYK colours: cyan (100,0,0,0); light orange (0,40,100,0); orange (0,60,100,0); red (0,100,100,0); pink (0,39,3,0); yellow (0,0,100,0); and beige (2,14,38,0). Finally, the last feature applies a random indentation to each text paragraph.

*2.2.7 Cover design*

The last step is the design of the book cover. We developed a method that generates simple typographic compositions with the title of the book (aligned to the top margin) and its author (aligned to the bottom margin) in uppercase letters and in the same font used in the text. The back cover includes information about the computer system that generated the book. The cover background colour is randomly selected from a set of predefined colours. The book title and author(s) need to be defined in the input document using a specific paragraph style. Alternatively, when this information is not specified, the system automatically sets the title to the first sentence of the paragraph composed in the biggest text size in the input document.

Experimental features also can be developed and applied to the covers. For instance, the presented version of the system includes experimental features that typeset the cover title in the largest font size possible.

# 4. Experimentation and Discussion

We performed an experiment to assess if, and to what extent, the proposed system is able to automatically design book layouts with different purposes and styles. In particular, we are interested in studying the ability of the system to perform two design tasks: (i) create books with distinguishable layout designs, *i.e.* to generate a set of books that present varied visual characteristics; and (ii) create visually coherent books, *i.e.* to generate a collection of books that follow and share the same visual style between them. Therefore, we conducted a survey to assess the visual diversity and coherence of designs created with the proposed system.

The following subsections describe our experimentation process. First, we explain the conducted experiment. Then, we report and discuss the obtained results.

**3.1 Experimental Method**

For the evaluation of visual diversity, we performed the following actions. First, we selected one public domain book. Then, we input this content into the system and generated 15 books while not manually setting any visual or typography attribute, *i.e.* the attributes were defined at random by the system within the predefined ranges based on the typographic styles, rules and principles defined by default in the system (see Table 1). Lastly, we presented the 15 generated book designs to a group of 42 participants and asked them to assess the layout diversity and/or coherence of the set.

The selected content was the book "Contos" written by the Portuguese author Eça de Queiróz and published in 1992.[2] This book comprises thirteen short stories, each one structured as a chapter, composed of about 73.330 words.

For the evaluation of visual coherence, we proceeded as follows. First, we selected a book from the set of 15 generated earlier to evaluate diversity. Then, we exported to file the settings used by the system to generate the selected book. Table 3 overviews the visual and typographic features of the chosen book design. Next, we input these settings into the system and created book designs for 5 different contents. Lastly, we presented the 5 generated designs to the same group of participants and asked them to evaluate the layout diversity and/or coherence of this set.

**Table 3.** System settings employed to generate the book designs used in the second part of the experiment, where their visual coherence is evaluated

| | |
|---|---|
| **Page size** (width × height) | 130 mm × 200 mm |
| **Page margins** (top \| inside \| bottom \| outside) | 12 mm \| 12 mm \| 13.7 mm \| 22 mm |
| **Running header and page numbering position** | Top page margin |
| **Grid** (columns number \| gutter size) | 1 \| n/a |
| **Title** (typeface \| size \| leading) | La Nord (Raoul Gottschling, 2006) \| 24 pt \| 27 pt |
| **Title alignment** (text box alignment on page \| text alignment) | Top page margin \| centre |
| **Body text** (typeface \| font size \| leading) | Antwerp (A2 Type, 2011) \| 10 pt \| 13 pt |
| **Body text alignment** (text alignment \| hyphenisation) | justify \| hyphenation on |
| **Cover colour** | CMYK (2, 14, 38, 0) |
| **Experimental features** | none |

The two sets of books were evaluated by the same group of testing participants through a survey. First, we presented to the participants a set of 15 designs generated at random and then a new set of 5 designs generated using the settings of one design selected from the first set. After observing each set of designs, each participant was asked to classify its layout diversity and/or coherence on a scale between 1 (very coherent) and 5 (very diverse). As already mentioned, the testing group includes 42 individuals. The age of the participants ranged from 19 to 49 years old.

---

[2] The book "Contos" by Eça de Queirós was retrieved from the Project Gutenberg. One may download the book at the following address www.gutenberg.org/ebooks/31347 (visited: 26 July 2022).

**3.2 Results and Discussion**

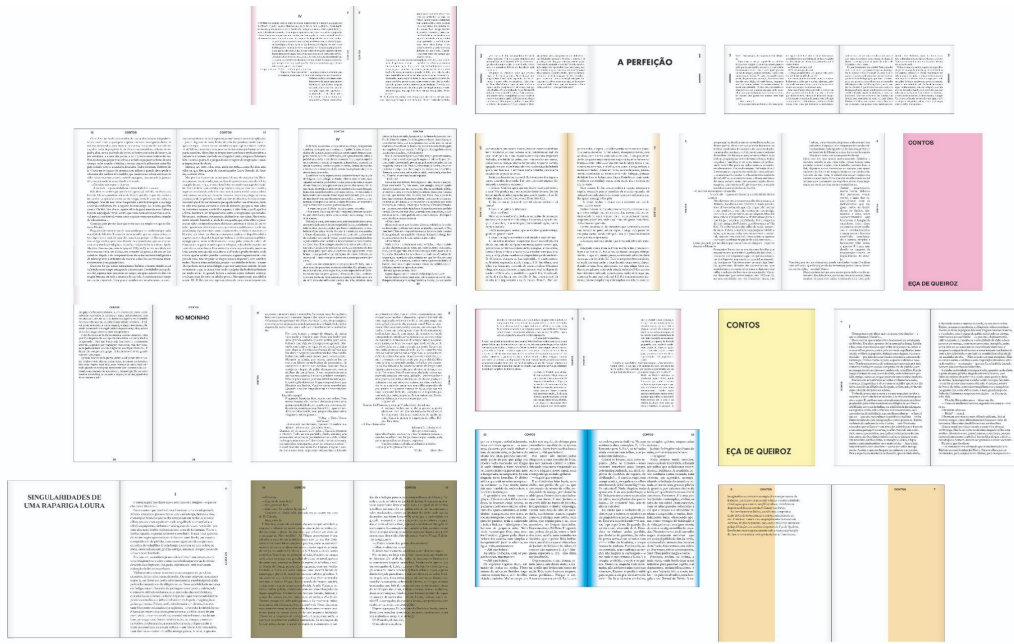

**Figure 3.** Examples of pages from books designed by the system using random settings and used in the first part of the experiment to assess their visual diversity. All designs generated by the system for this paper can be found in the supplementary files.

Figure 3 presents several pages composed automatically by the system for the first part of the experiment, where the visual settings are defined at random. Looking at the resulting designs, we noticed that most of the designs (12 out of 15) are portrait-oriented. This is due to the automatic classification of the input content by the system as a long reading book and therefore the probability of selecting a portrait format is higher. Nevertheless, the format of the generated books exhibits slight variations between them. Concerning the text box, the grids and the typefaces used, we noticed the influence of the typographic principles and rules encoded in the system. Most designs present the body text typeset in the justified text (13 out of 15) over a one-column grid (12 out of 15), which are characteristics considered suitable for long reading books. Nevertheless, it is possible to observe diversity in other aspects. The dimensions of the margins and the size of the grid gutter, when it exists, vary between books. The used font, text size and leading and position of the running and page numbers also change. Additionally, it is possible to observe that different experimental features were used, alone or combined, in 6 of the generated designs.

Observing books created when the system settings are imported from a previously generated book, we can notice the share of visual characteristics among the resulting designs (*e.g.* book format and size, page margins, grid or used typefaces) and their similarity to the initial design. Figure 4 shows pages from the book that sourced the settings file (Figure 4a), which was selected from the set used in the first part of the experiment (Figure 3), along with pages of some books used in the second part of the experiment (Figure 4b). One should note that Figure 3 and Figure 4 depict only a small portion of the designs generated for the experiment, which can be all consulted in the supplementary files.

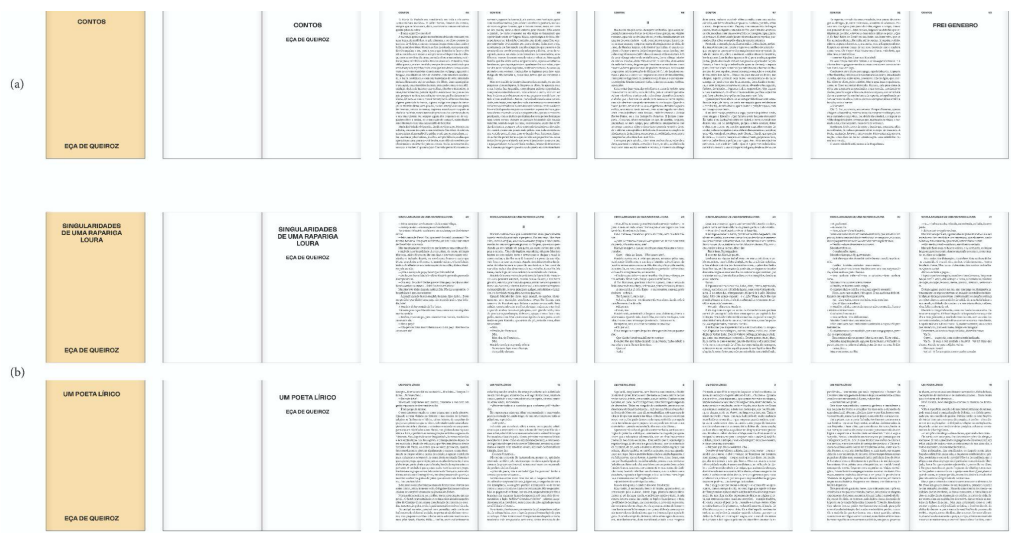

**Figure 4.** Examples of pages from books designed by the system using the same settings. The top row of pages (a) belongs to a book selected from the set used in the first part of the experiment. The other two rows of pages (b) belong to two books of the set used in the second part of the experiment, generated using the settings sourced by the book selected from the first set (a). All designs generated by the system for this paper can be found in the supplementary files.

The conducted survey indicates that the system is able to generate both diverse and cohesive book designs. The chart in Figure 5 shows the distribution of answers obtained in the survey.

When we asked the survey participants to classify the diversity and/or coherence of the first set of books, which were generated at random by the system, the majority of participants (35 out of 42) considered them as diverse (19 participants) or very diverse (16 participants). Only a few participants considered that this set of books was very cohesive (2 participants) or cohesive (5 participants). When participants performed the same task for the second set of books, which were generated from a specific settings file, the majority of participants (38 out of 42) considered that the generated designs were visually coherent between them. Most of them (30 participants) considered that this second set of designs present a high-level coherence among them.

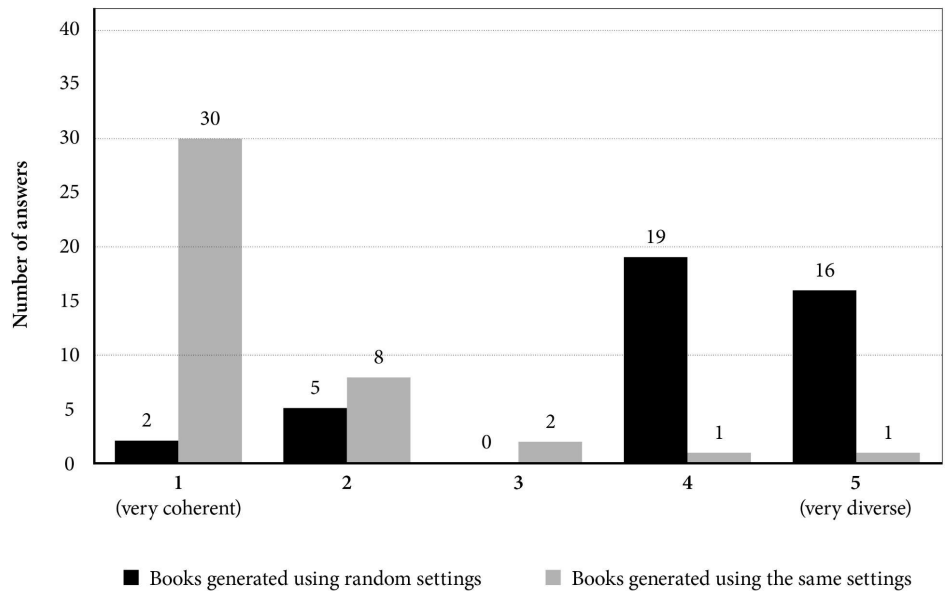

**Figure 5.** Distribution of the answers obtained in the user survey conducted to evaluate the visual coherency and diversity of different books generated by the system. Black bars regard user evaluations of books created using random settings chosen automatically by the system. Grey bars regard user evaluations of books created using the same settings imported from a selected settings file.

The survey results revealed that variables and properties defined by the system can create visually diverse layouts. Although the system engine determines the features of books based on a set of predefined typographic rules and principles, it also employs probabilistic mechanisms to define some attributes. This allows the occasional definition of attributes in an unexpected manner, promoting visual variation on the resulting layouts. This is visible, for instance, in the first part of the experiment, where books are created with the different formats even though the system tends to avoid the use of landscape format in long reading.

The obtained results expose the exploratory nature of the proposed system, which demonstrates high potential to stimulate and foster graphic designers' creativity and experimentation in the different stages of the design process. The system presents itself as a co-creativity tool which enables editorial designers to explore multiple conceptual and visual possibilities in an accessible, easy, and effortless manner. This is possible by enabling users to not only define values of the different attributes but also by enabling them to define the level of autonomy of the system and/or target the exploration of certain properties. Therefore, the system can be used in most stages of the design process, from the earlier and exploratory stages (when designers can take advantage of random generation to look for new types of layouts) to the final stages (when designers need to fine-tune one or more graphic attributes).

The system allows exporting data to settings files that encode the design of the book and, later, can be loaded into the system to create other books which are visually similar. As demonstrated by the survey results, the proposed system can automatically create highly coherent designs when the book properties and characteristics are prior determined and stored in these settings files. Users can also include in the system their desired book properties both directly on the system interface and/or by modifying the typographic principles and rules used by the system. The proposed system is also a useful tool to automate the design of books that need to follow a set of restricted typographic and visual attributes (*e.g.* when it is necessary to design new books that will be part of a collection or a series). Thus, besides its exploratory nature, the system also enables the automatisation of some editorial design tasks and routines.

With the proposed system, users do not require other software tools to manipulate and produce the resulting book designs since it operates inside of the popular editorial design software Adobe InDesign and the generated books are made available as editable documents. This way, it is fully integrated into the typical working environment of editorial designers. This allows its use both as an exploratory and as an automation tool, thus empowering designers to easily edit or fine-tune the output designs directly in a familiar environment.

The experimental features implemented by the system can be observed in some of the generated books and definitely contribute to their diversity and variation. Nevertheless, we noted that some generated designs exhibit certain graphic limitations. This may be related to the fact that the generated designs comply with the same default typographic rules and principles. Although users can manipulate these rules, this will primarily change the typeset of the book but it will not include new visual features. For this reason, the system facilitates the addition of new features as well as their control. In this sense, one can add new features to the system for exploration and automatisation purposes. We believe that this possibility may allow the system to solve some lack of distinguishable visual features of the output, including in the design of the covers.

In summary, the experimental results demonstrate the ability of the proposed system to automatically generate finished and functional designs from scratch. Furthermore, the results reveal the potential of the system as a useful exploratory tool in the context of book typesetting and editorial design. It may be operated by graphic designers when they are searching and exploring new conceptual and visual perspectives, fine-tune a book characteristic, and/or design books that must be coherent with a given set of typographic rules. In addition, we consider that the automation provided by the system has great potential in varied graphic design commercial scenarios, *e.g.* the automatic design and production of customised books, which is relevant for print-on-demand applications, or the effortless typeset of books for an existing one-book collection.

## 5. Conclusion

We have presented a novel approach to computationally design books. The presented system implements a generative design process which takes advantage of the scripting capabilities of Adobe InDesign to procedurally typeset books from content provided by the user. We have shown the ability of the system to (i) create book designs that consistently comply with a series of typographic rules, styles and principles identified in the literature; (ii) produce visually diversified books from the same input content; and (iii) produce visually coherent books with different contents.

The work presented in the paper may challenge the typical roles of both the tool and the designer. First, by automatically creating and suggesting design alternatives, the tool ends up playing a more active role in the design process. Then, by modifying and developing custom tools, the designer is no longer a mere tool user and becomes the author of tools tailored to specific needs. We believe this shift can be fruitful since it enables the exploration and discovery of new technical and creative possibilities.

This work can hopefully provide directions to further research on generative processes for supporting design exploration and finding unique designs. In the particular case of typography, generative approaches such as the one presented in the paper can be useful and reveal great potential, especially in the current print-on-demand market and digital publishing, where each publication may be unique.

Our future work will move in the direction of employing Artificial Intelligence techniques, such as Evolutionary Computation and Machine Learning, to enable a deeper exploration of the vast space of book designs that can be achieved with the system and also to automatically suggest settings to designers according to their needs or goals.

## 6. Acknowledgments

We would like to express our gratitude to all the participants in the evaluation sessions. This work is partially supported by the Foundation for Science and Technology, I.P./MCTES (Portugal) through national funds (PIDDAC), within the scope of project UIDB/00326/2020 or project code UIDP/00326/2020. Sérgio M. Rebelo was funded by FCT under the grant SFRH/BD/132728/2017 and COVID/BD/151969/2021.